\documentclass[a4paper, 12pt, titlepage]{article}
\usepackage{fullpage}
\usepackage{setspace}
\usepackage[dvips]{graphicx}
\usepackage{amssymb}
\usepackage{fullpage}
\usepackage{subfigure}
\begin{document}

\begin{center}
\large{\textbf{Time-Resolved Ring Structure of Backscattered Circularly \\ Polarized Beams from Forward Scattering
Media
}}
\end{center}

\begin{center}
Kevin G. Phillips, M. Xu, S.K. Gayen, R.R. Alfano
\end{center}

\begin{center}
\textit{Institute for Ultrafast Spectroscopy and Lasers, Department of Physics,\\ 
City College of New York, 
Graduate School of the City University of New York,\\ 
New York, New York 10031}
\end{center}
\noindent
\begin{center}
\textit{Abstract}
\end{center}
The backscattering of circularly polarized light at normal incidence to a
half-space of scattering particles is studied using the Electric Field
Monte Carlo (EMC) method.
The spatial distribution of the backscattered light intensity is examined for both the time-resolved
and continuous
wave cases for large particles with anisotropy factor, $g$, in the range 0.8
to 0.97.
For the time-resolved case,
the backscattered light
with the same helicity as that of the incident beam (co-polarized) is found to form a ring centered on the
point of incidence. The ring expands and simultaneously grows weak as time increases.
The intensity of backscattered light with helicity opposite
to that of the incident beam (cross-polarized) is found to exhibit a ring behavior for
$g \geq 0.85$, with significant backscattering at the point of incidence.
For the continuous-wave case no such ring pattern is observed in backscattered light for either helicity.
The present EMC study suggests
that the ring behavior can only be observed in the time domain,
in contrast to previous studies of light backscattered from forward scattering media based on the scalar time-independent
Fokker-Planck approximation to the radiative transfer equation.
The time-dependent ring structure of backscattered light may have potential use in
subsurface imaging applications.

\pagebreak
\section{Introduction}

Polarized light
backscattered from highly scattering media carries
medium-specific information that can be utilized in polarization sensitive imaging
techniques\cite{gilbert}-\cite{ni}.
How light depolarizes and propagates from and within
a scattering medium remains an active topic of research
in optical imaging of targets embedded in highly scattering media.
Polarization sensitive imaging
techniques depend on the
level of anisotropy of the scattering medium, which in turn
depends on the relative size of scattering particle and wavelength of light used.
When the particle diameter,
$a$, is small compared to
the wavelength, $\lambda$, of incident light, the transport mean free path, $l_{t}$, becomes equal to the
scattering mean free path, $\l_s$, and isotropic scattering results. In isotropic scattering linearly polarized light
has a longer depolarization length than circularly polarized light and tends to backscatter with the same polarization as
the input whereas circularly polarized light backscatters
with its helicity flipped. This is attributed to single scattering events with large scattering angle.
In the case of
anisotropic scattering for large particles, where $a \geq \frac{\lambda}{4}$ yielding $l_{t} > l_s$, circularly polarized light has a longer
depolarization length than linearly polarized light and
backscatters with preserved helicity\cite{xu1}. This process is termed polarization memory\cite{MacKintosh}.
In this case linearly
polarized light is backscattered with some elliptic polarization. This is attributed to forward scattering
within the medium, causing the polarization of linearly polarized light to become randomized over shorter
distances than the
circularly polarized light.

The questions that need to be addressed include the following.
Is light backscattered primarily from the point of incidence?
Does backscattered light take special pathways inside the media towards the incident surface?
Does the backscattered light exhibit time-dependent features?
These questions warrant study as it is strongly
anisotropic scattering
in the forward direction that is encountered
in biomedical optical imaging.
Such knowledge
would give an indication of the scattering behavior of  light as it travels $\textit{inside}$
the scattering medium.

An analysis of the spatial dependence of backscattered circularly polarized beams was first carried out
by Kim and Moscoso\cite{kim1} who used the scalar time-independent Fokker-Planck equation, an approximation
to the radiative transfer equation.
They predict a time-independent
circularly symmetric peak centered on the point of incidence, henceforth referred to as a ``ring'' or
``ring-peak'',
in the backscattered light and postulate
that light comprising the ring is of the same helicity as the circularly polarized incident beam.

In this study, the Electric Field
Monte Carlo method\cite{xu} is used to investigate the spatial distribution of time-dependent (time-resolved),
in addition to steady state (continuous wave), properties of light backscattered when a pencil-like beam is
incident on a half-space of
forward scattering particles whose
anisotropy factors range from $g = 0.8$ to $g = 0.97$. The EMC program uses the exact phase function
of the individual scatterers to allow for a more physical simulation of light propagation. The EMC simulation
results suggest
that the backscattered light
forms rings that can only be observed in the time domain and can be seen in both
the co-polarized and cross-polarized
backscattered light in the presence of
forward-peaked scattering with $g \geq 0.85$.

\pagebreak
\section{The Electric Field Monte Carlo Method}

Polarized light propagation in
turbid media, including human tissues\cite{akira}, may be described by the time-dependent vector radiative transfer
equation (RTE)\cite{chandra}:
\begin{equation}
\begin{array}{ll}
 (c_o^{-1}\partial_t + \hat{s}\cdot\nabla_r + \sigma_t) \mathbf{I}(\vec{r},\hat{s},t) & =
\int_{\Omega} \mathbf{P}(\hat{s} \cdot \hat{s}') \cdot \mathbf{I}(\vec{r}, \hat{s}', t)ds'
\end{array}
\end{equation}
\noindent
\noindent
Here $\mathbf{I}$ is the Stokes vector, defined by $\mathbf{I} = (I~,Q~,U~,V)^T$, where
T denotes transpose. The components of the Stokes vector are defined in terms of the
two orthogonal complex electric field components, $E_{1,2}$, perpendicular to the direction
of the incident light:
$I = \left< E_1E_1^* + E_2E_2^* \right>$, $Q = \left< E_1E_1^* - E_2E_2^* \right>$,
$U = \left< E_1E_2^* + E_2E_1^* \right>$ and $V = i\left< E_1E_2^* - E_2E_1^*\right>$.
 $\left< A \right>$ denotes time averaging of the quantity $A$,
\noindent
$c_o$ is the speed of light in the medium, $\sigma_t$ is the total
extinction cross section and $\Omega$ denotes the unit sphere. $\mathbf{P}$ denotes the $4 \times 4$ phase matrix.

Due to the current absence of analytical solutions to the RTE
within a bounded medium, numerical studies of polarized light propagation involving Monte Carlo
simulation and numerical solutions to the RTE are often the
most available tools to explore polarized light propagation in turbid media from a theoretical perspective.
These methods\cite{xu},\cite{kim4} - \cite{tynes}  have been extensively used to characterize different types of
scattering media such as
particle suspensions, and biological materials.

The Electric Field Monte Carlo method (EMC) traces the multiply scattered electric
field of incident light in time to simulate the time-resolved
propagation
of polarized light in turbid media.
At each step in time, the EMC program simulates light scattering, absorption, or unhindered propagation
giving the scattered electric field, $\mathbf{E}'(t)$.
The parallel and perpendicular components of the complex
electric field
are updated  with respect to the instantaneous scattering plane by the amplitude scattering matrix and simultaneous
rotations of
the local coordinate system spanned by the components of the electric field and the propagation direction. As described
in reference\cite{xu}, if $\mathbf{m}$ and $\mathbf{n}$ are unit vectors in the directions of the parallel
and perpendicular
components of the electric field with respect to the previous scattering plane and $\mathbf{s}$ is the electric field
propagation direction prior to the current scattering event then the propagation direction $\mathbf{s}'$ of the electric
field after the current scattering event is given by:\\
\begin{equation}
\mathbf{s}' = \mathbf{m}\sin(\theta)\cos(\phi)+\mathbf{n}\sin(\theta)\sin(\phi)+\mathbf{s}\cos(\theta)
\end{equation}

where $\theta$ is the scattering angle and $\phi$ is the azimuthal angle of the current scattering. The current
scattering plane is spanned by $\mathbf{s}$ and $\mathbf{s}'$. The unit vectors in the direction of the parallel
and perpendicular electric field components with respect to $\mathbf{s}$ and $\mathbf{s}'$ are given prior to scattering by:
\begin{equation}
\begin{array}{ll}
\mathbf{e}_1 & =~\mathbf{m}\cos(\phi)~+~\mathbf{n}\sin(\phi)\\
\mathbf{e}_2 & =-\mathbf{m}\sin(\phi)~+~\mathbf{n}\cos(\phi)
\end{array}
\end{equation}
and after scattering are given by:

\begin{equation}
\begin{array}{ll}
\mathbf{e}_1' & =~\mathbf{m}\cos(\theta)\cos(\phi)+\mathbf{n}\cos(\theta)\sin(\phi)-\mathbf{s}\sin(\theta) = \mathbf{m}'\\
\mathbf{e}_2' & =~\mathbf{e}_2 = \mathbf{n}'
\end{array}
\end{equation}

The local coordinate system $(\mathbf{m},\mathbf{n},\mathbf{s})$ is rotated to
($\mathbf{m}'$,$\mathbf{n}'$,$\mathbf{s}'$) with the components of the incident electric field,
$\mathbf{E}$ = $E_1 \mathbf{m} + E_2 \mathbf{n}$,
being updated correspondingly by:
\begin{equation}
\mathbf{E}' = E_1'\mathbf{m}' + E_2'\mathbf{n}' = (S_2\mathbf{E}\cdot\mathbf{e}_1)\mathbf{m}'+(S_1\mathbf{E}\cdot\mathbf{e}_2)\mathbf{n}'
\end{equation}
where $S_2(\theta)$ and $S_1(\theta)$ are the diagonal elements, respectively, of the amplitude scattering matrix. The
amplitude scattering matrix is diagonal as a result of the symmetry of the spherical scatterers.

To summarize, the local coordinate system ($\mathbf{m},\mathbf{n},\mathbf{s}$) is updated to
($\mathbf{m}'$,$\mathbf{n}'$,$\mathbf{s}'$) by:

\begin{equation}
\left( \begin{array}{ll}
\mathbf{m}' \\
\mathbf{n}' \\
\mathbf{s}'
\end{array} \right)
=
\left( \begin{array}{lll}
\cos(\theta)\cos(\phi) & \cos(\theta)\sin(\phi) & -\sin(\theta) \\
-\sin(\phi) & ~~\cos(\phi)  & ~~~~0 \\
\sin(\theta)\cos(\phi) & \sin(\theta)\sin(\phi) & ~\cos(\theta)
\end{array} \right) \cdot
\left( \begin{array}{ll}
\mathbf{m} \\
\mathbf{n} \\
\mathbf{s}
\end{array} \right)
\end{equation}
\noindent
and the electric field is updated by:

\begin{equation}
\left( \begin{array}{ll}
E_1'\\
E_2'
\end{array} \right) = \frac{1}{\sqrt{F(\theta, \phi)}}
\left( \begin{array}{cc}
S_2\cos(\phi) & S_2\sin(\phi) \\
-S_1 \sin(\phi) & S_1\cos(\phi)
\end{array} \right) \cdot
\left( \begin{array}{ll}
E_1\\
E_2
\end{array} \right)
\end{equation}
\noindent
where the scattered wave intensity, $F(\theta, \phi)$,
into direction $\mathbf{s}'$ (characterized by $\theta, \phi$) has
been introduced
as a normalizing factor\cite{xu}. This normalization factor insures that the scattered light intensity is conserved at
each scattering event. Absorption of light is accounted for by adjusting the weight of the scattered photons
as in standard Monte Calro simulations\cite{rubin}.

The coordinate system and electric field are updated simulatneously
at each scattering event defined by the sampling of $\theta, \phi$ from the normalized phase function $p(\theta, \phi)
= F(\theta, \phi)/\pi Q_{sca} (ka)^2 $ where $Q_{sca}$ is the scattering efficiency, $k$ is the wave vector of
incident light on
a spherical particle and $a$ is the radius of the particle. To sample $\theta$
and $\phi$ the phase function of the particles is determined numerically at the beginning of the calculation.

After $n$ consecutive scattering events
the state of a photon is given by the local coordinate system
($\mathbf{m}^{(n)}$,$\mathbf{n}^{(n)}$,$\mathbf{s}^{(n)}$), the complex electric
field components $E_{1,2}^{(n)}$, the optical path length $l$ (distance travelled by photon in the medium)
and weight, $w$.
Initially unity at incidence, the
weight is multiplied by the albedo of the scatterer at each scattering event. Once the photon hits the detector
the electric field at the detector, $\mathbf{E}_d$, and Stokes vector, $\mathbf{I}$, is increased according
to the detected light increment:
$w^{1/2}\mathbf{E}_d = w^{1/2}(E_{1}^{(n)}\mathbf{m}^{(n)}+E_{2}^{(n)}\mathbf{n}^{(n)})exp(2\pi i l / \lambda )$
\cite{xu}.
\\
\section{Results}
A collimated beam, $N\delta(t)\delta(\vec{r})$, where $N = 2\times10^{11}$ circularly polarized photons,
is normally incident on
a slab of scatterers of length $25[l_s]$.
Mie calculations were used to determine $g$, $l_s$,
$l_{t}$ and the phase function $p(\hat{s} \cdot \hat{s}')$ prior to running the EMC program. To optimize computation
time the calculation was broken up into parts involving a reduced number of photons and run simulataneously
on the CUNY-GC research cluster.
A $50[l_s] \times 50[l_s]$ detecting area with a resolution of 0.1$[l_s]$ was used to monitor the
light backscattered around the incident direction
over a time range of 50[$\frac{l_s}{c}$], with a time resolution of 0.1$[\frac{l_s}{c}]$.
Two types of collection schemes were explored. Firstly one in which backscattered light with arbitrary propagation
direction was collected and the second collected only light exactly backscattered
(propagation direction anti-parallel to the
incident beam).
Computation time was further reduced
by employing the circular symmetry of the spatial profiles of the backscattered light to plot the
backscattered intensity as a function of radial distance from the point of incidence. This circular symmetry is
attributed to the normal incidence of the beam. All results presented
correspond to the collection of backscattered light with arbitrary propagation direction. As well each plot
of backscattered light intensities, weather spatial or radial, has been normalized according to
the total number of photons,$N = 2\times10^{11}$,  used in the simulation.
\\
\subsection{Time-Resolved Backscattering}
For time-resolved detection of light backscattered with the same polarization as the incident beam
(co-polarized backscatter),
a ring centered on the point of incidence of the beam is found to form. Initially the radius of the ring expands
linearly with time. The intensity of light at the ring-peak grows weaker as the ring expands.
Fig. (1) shows the intensity of the co-polarized backscattered beam for $g = 0.96$
at different times. Ring formation corresponds
to the red peak which noticeably
expands and becomes weaker (as seen in the change in colorbar scales) as time progresses.
The backscatted intensity at the ring-peak is more pronounced than backscattering
at the point of
incidence (the center of the plots).
\begin{figure}
\centering
\includegraphics[scale = 0.75]{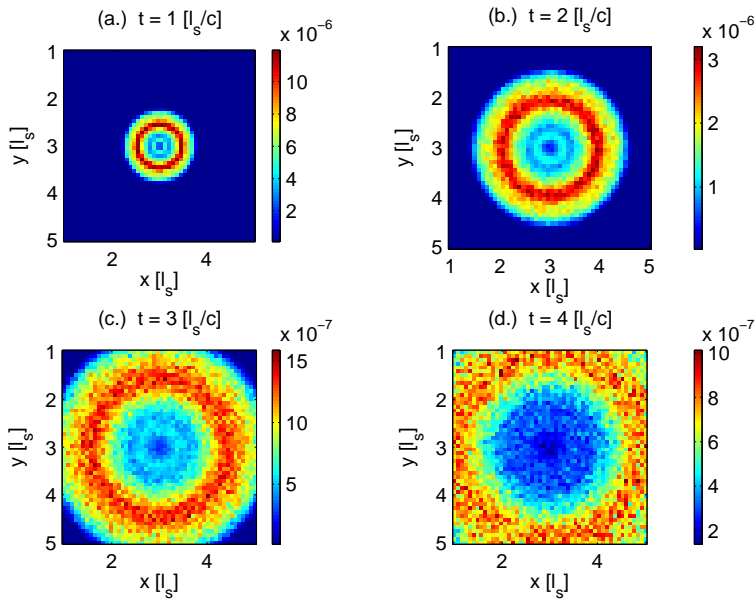}
\caption{Backscattered time-resolved intensity of co-polarized light for $g = 0.96$
at times: (a.) t = 1$[\frac{l_s}{c}]$, (b.) t = 2$[\frac{l_s}{c}]$,
(c.) t = 3$[\frac{l_s}{c}]$ and (d.) t = 4$[\frac{l_s}{c}].$}
\end{figure}
Ring formation was observed in co-polarized backscattered light for all values of
$g$ ranging from 0.8 to 0.97.

Fig. (2) shows radial profiles of the backscattered intensity with same helicity
as the incident beam
for $g$ = 0.8, 0.85, 0.9 and 0.95 at a single time $t = 2 [\frac{\l_s}{c}]$. An important feature
illustrated in Fig. (2) is the weakening of the ring-peak as $g$ increases. This is due
to increased forward-peaked scattering within the medium and a resulting
decrease in the likelihood of return of the photons with polarization of the incident beam.
\begin{figure}
\centering
\includegraphics[scale = 0.75]{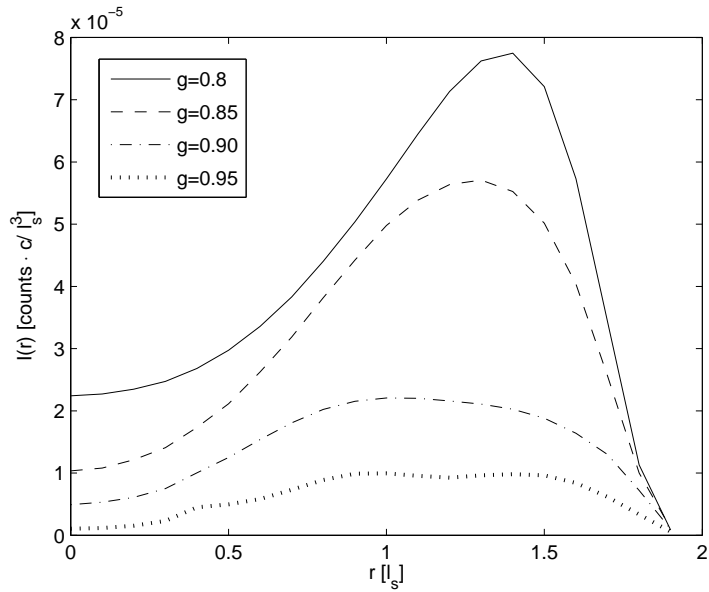}
\caption{Comparison of backscattered intensity of light co-polarized with the incident beam for different
anisotropy factors at t = 2$[\frac{l_s}{c}]$.}
\end{figure}

For each value of $g$, the ring behavior in the backscattered co-polarized light
evolved in a similar way:
initially the ring expanded linearly in time away from the point of incidence, growing weaker at each time step,
until it reached a maximal
radius at which point the ring-peak flattens and
backscattering at the point of incidence begins to take over, see Fig. (3).
\begin{figure}
\centering
\includegraphics[scale = 0.75]{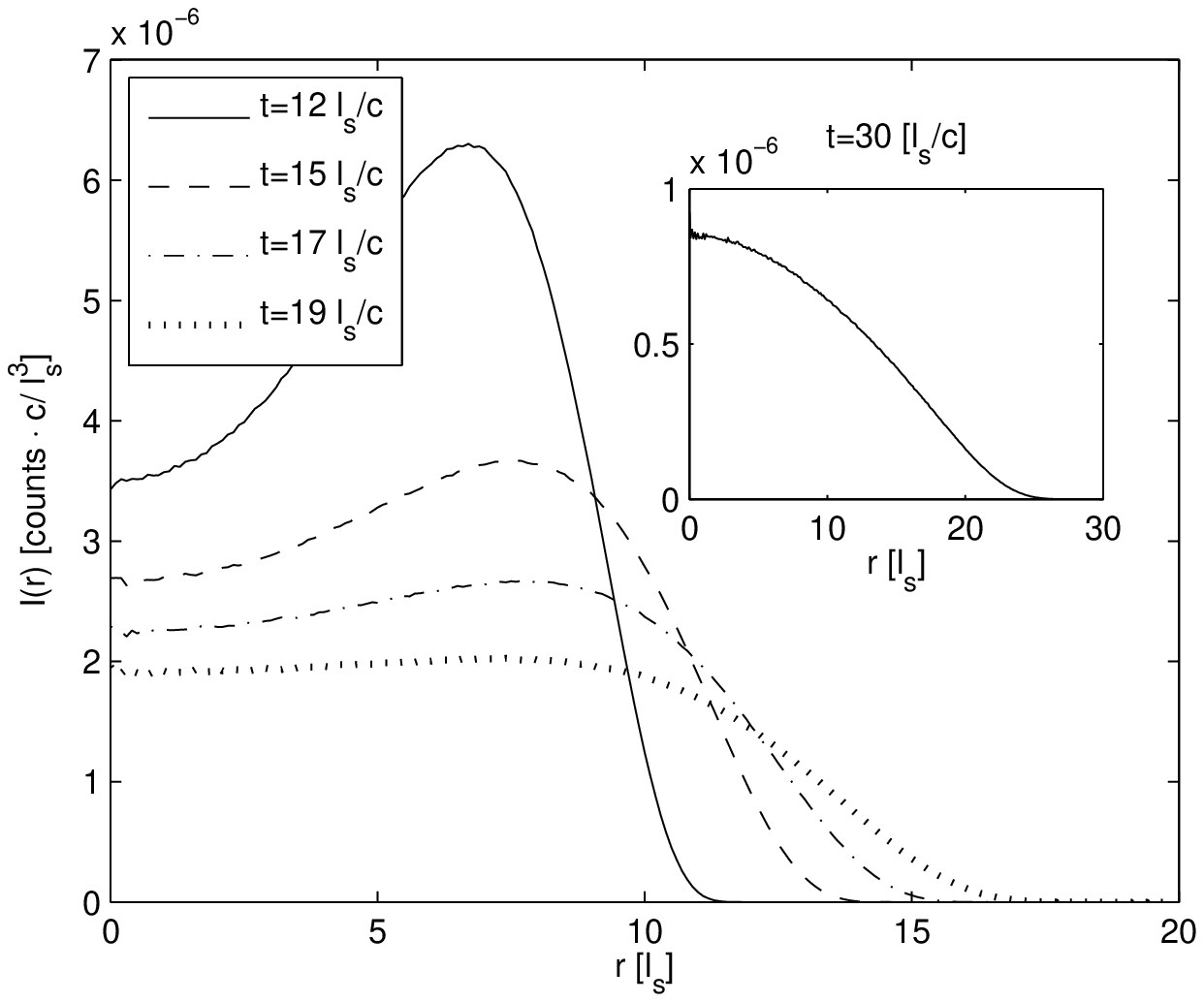}
\caption{Time evolution of co-polarized backscattered light for $g = 0.80$.
Note at t = 19$[\frac{\l_s}{c}]$  plateauing of the ring-peak occurs. $\textit{Inset:}$ the radial profile at t = 30$[\frac{l_s}{c}]$, convergence to a Guassian-like distribution.}
\end{figure}
After sufficient time has passed
the backscattered light converges to a Gaussian-like distribution with peak at the point of incidence.
The maximum radius of the ring and time over which the ring is present (ring-lifetime),
increases
for increasing values of $g$, prior to this flattening effect.
This is a consequence of increased forward scattering as $g$ increases.

The ring behavior occurs due to successive near-forward scattering events which preserve
the polarization state
of the photon as a result of polarization memory effects\cite{xu1},\cite{MacKintosh} described earlier.
The successive near-forward scattering
events give rise to arc-like trajectories
of photons as they travel into the medium. Rings with smaller radii are composed of
shallow penetrating photons
while rings with larger radii are composed of
photons that travel deeper,
giving depth information about the scattering medium.

As $g$ increases, the number of near-forward scattering events
needed to bring a photon back to the surface increases, giving rise to weaker ring-peaks and prolonged ring life-times.
The time-dependence of the rings is a result of photons penetrating deeper into the medium as a result of forward
scattering: photons that travel further into the media take longer to backscatter along the semi-circular trajectories
giving rise to ring formation at later times.
Fig. (4) shows a schematic depiction of this process.
\begin{figure}
\centering
\includegraphics[scale = 0.75]{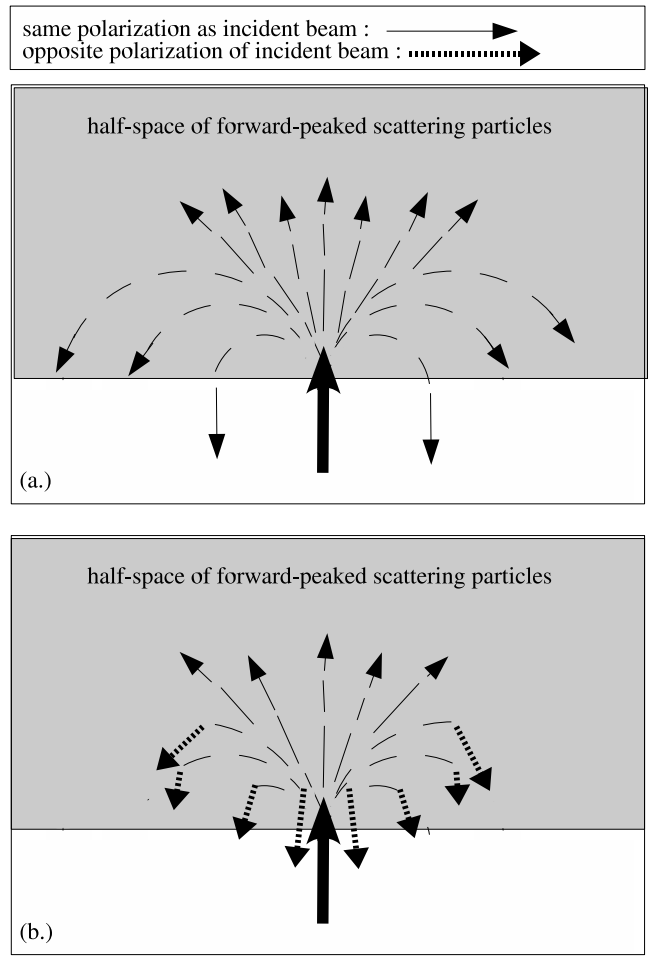}
\caption{Schematic diagram illustrating light pathways contributing to ring formation with
bacscattered light (a.) co-polarized , and
(b.) cross-polarized, with the incident beam.}
\end{figure}
Fig. (5) shows the linear dependence on time the ring radius displays
up to the point of plateauing of the ring peak. Beyond this plateauing the radial position of the peak of
light intensity
moves to $r_{max} = 0$ as backscattering at the point of incidence takes over. This feature can be
seen in the ``fall'' back to zero of each curve in Fig. (5).
As well, it illustrates the relationship between anisotropy and the duration of the ring behavior of the backscattered
light: the ring is prevalent for longer times as $g$ increases.
\begin{figure}
\centering
\includegraphics[scale = 0.75]{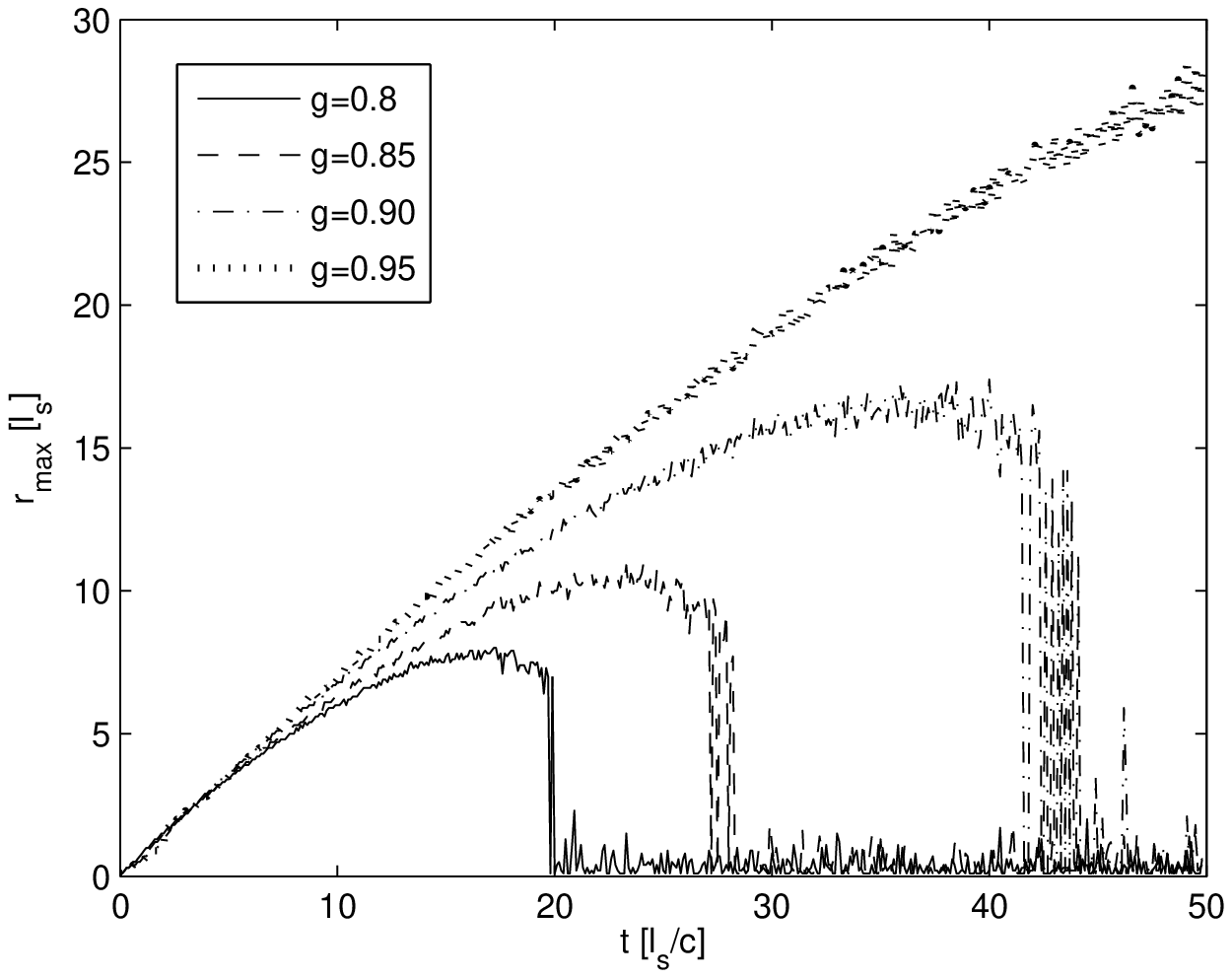}
\caption{Time dependence of the radius of the ring-peak of co-polarized backscattered light
for various values of $g$.}
\end{figure}

Light backscattered with the opposite helicity of the incident circularly polarized beam is found to display a different
ring-like
behavior for sufficiently strong forward-peaked scattering, $g \geq 0.85$.
Light is primarily backscattered at the point of incidence, as a result of large angle scattering,
with a secondary ring-peak forming some distance away from
the point of incidence.
As in the case of the helicity preserved
backscattered light, the ring radius increases in time and the peak associated with the ring decreases simultaneously.

Fig. (6) displays the spatially resolved backscattering
of light of opposite helicity with respect to the incident beam (cross-polarized backscatter) for $g=0.96$.
The dark red center of the
plots correpsonds to backscattering at the point of incidence.
The secondary light blue peak is the ring-peak. Light backscattered at the point of incidence results from the finite
tail of the phase function, $p(\hat{s} \cdot \hat{s}')$, peaked about the scattering angle $\hat{s}\cdot\hat{s}'=\pi$.
Because most of
the light is transmitted as
a result of forward-peaked scattering, large angle scattering is dominant when looking at the backscattered light. Large
angle scattering is also responsible for the helicity flip of the backscattered light\cite{MacKintosh}.
\begin{figure}
\centering
\includegraphics[scale = 0.75]{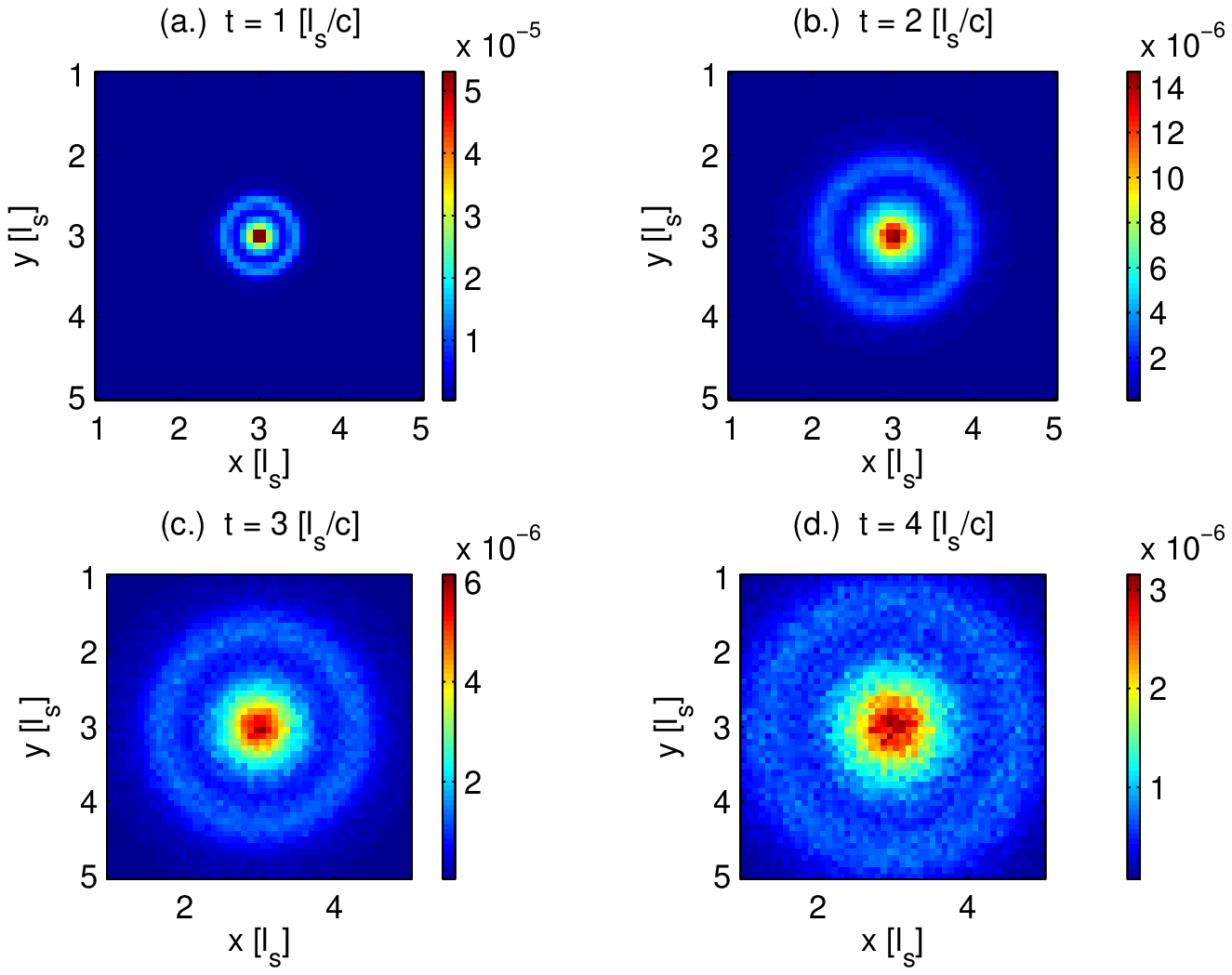}
\caption{Time-resolved intensity of cross-polarized backscattered light for $g = 0.96$
at times (a.) t = 1$[\frac{l_s}{c}]$, (b.) t = 2$[\frac{l_s}{c}]$, (c.) t = 3$[\frac{l_s}{c}]$ and (d.) t =
4$[\frac{l_s}{c}$]. }
\end{figure}

As in the case of polarization preserved backscattering, the ring observed for light bacskcsattered with opposite
polarization of the incident beam moves away from the point
of incidence as time progresses, growing weaker as it does so, until a plateauing occurs with eventual
convergence to a Gaussian-like distribution.
Fig. (7) displays the plateuing for the case
in which $g = 0.85$. The inset displays the gaussian-like convergence, with peak at the point of incidence,
for long times. Again, the ring behavior is attributed to successive near-forward scattering events with the
presence of a small number (mostly, single) of large-angle scattering events which are responsible for the
polarization flip of
the backscattered light, see Fig. (4). Light that travels deeper into the medium is responsible for ring formation at later
times while light that travels to smaller depths is responsible for ring formation at early times. As $g$ increases
the ring-peak deminishes due to increased forward scattering resulting in greater transmittance of light, see
Fig. (8).

\begin{figure}
\centering
\includegraphics[scale = 0.75]{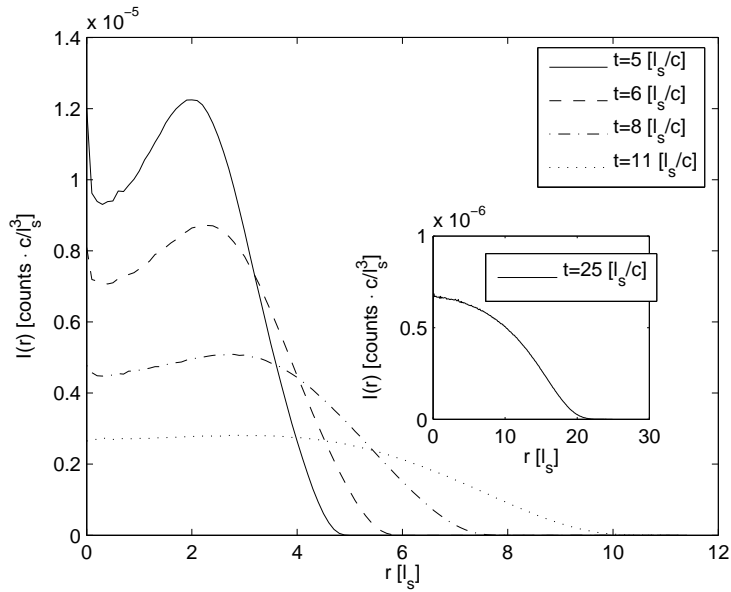}
\caption{Time evolution of cross-polarized backscattered light for $g = 0.85$.
Note at t = 11$[\frac{l_s}{c}]$ plateauing of the ring-peak occurs. Inset:
the radial profile at t = 25$[\frac{l_s}{c}]$, convergence to a Gaussian-like distribution.}
\end{figure}

\begin{figure}
\centering
\includegraphics[scale = 0.75]{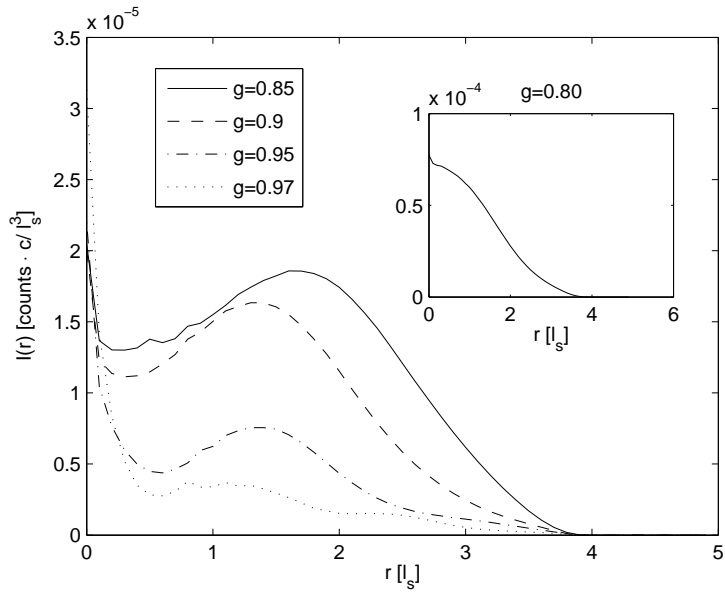}
\caption{Comparison of cross-polarized backscattered light for various anisotropies, t = 4$[\frac{l_s}{c}$]
.}
\end{figure}
Lastly, a comparison of rings formed by co-polarized and cross-polarized backscattered light
at equal times and equal anisotropies
reveals that the radial position of ring-peaks composed of light with preserved helicity
are greater than the radial position of ring-peaks composed of light with flipped helicity, see Fig. (9).
This is attributed to large angle scattering in which photons deviate from the successive near-forward
scattering paths comprising the arc-like trajectories going back towards the incident surface.
As a result, the photons that undergo large angle scattering towards the incident surface
have their polarizations flipped and traverse a shorter path inside the scattering medium,
resulting in a smaller ring radius, Fig. (4).
\begin{figure}
\centering
\includegraphics[scale = 0.75]{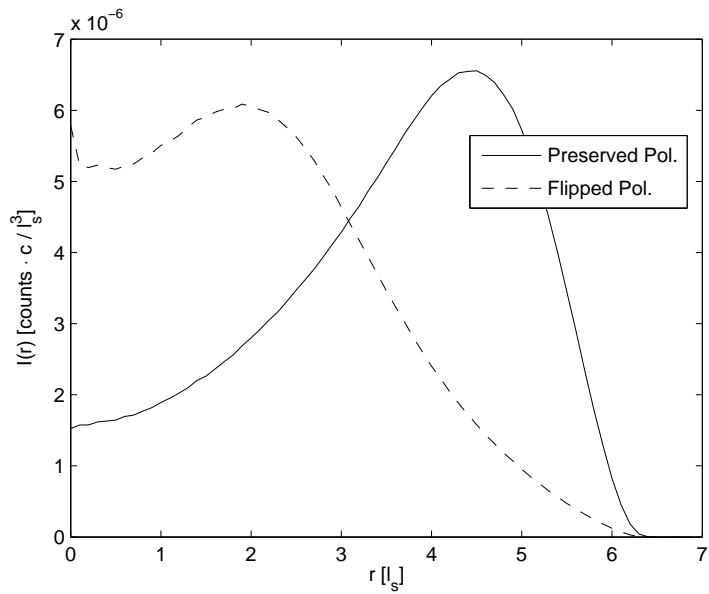}
\caption{Comparison of backscattered light with opposite and perserved polarization at t = 6.5$[\frac{l_s}{c}$] and with $g= 0.90$.}
\end{figure}

\subsection{Continuous-Wave Backscattering}

In the continuous-wave case, no ring formation is observed in the backscattered light of either
preserved or flipped helicity with respect to the incident beam. Fig. (10) shows the spatially resolved
continuous-wave backscattered intensity of both helicities with $g=0.96$.

\begin{figure}
\centering
\subfigure[]{\includegraphics[scale = 0.5]{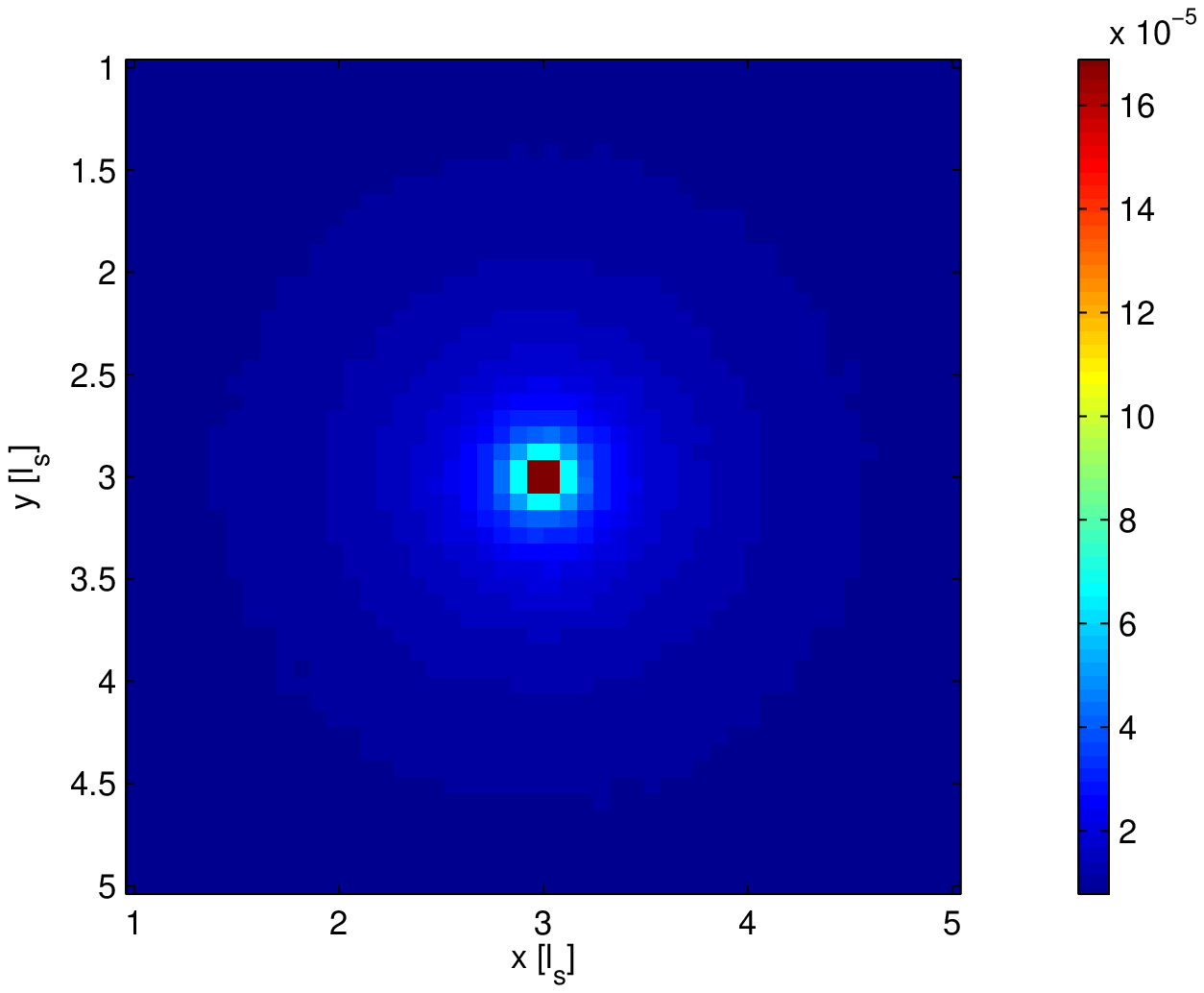}}
\hfill
\subfigure[]{\includegraphics[scale = 0.5]{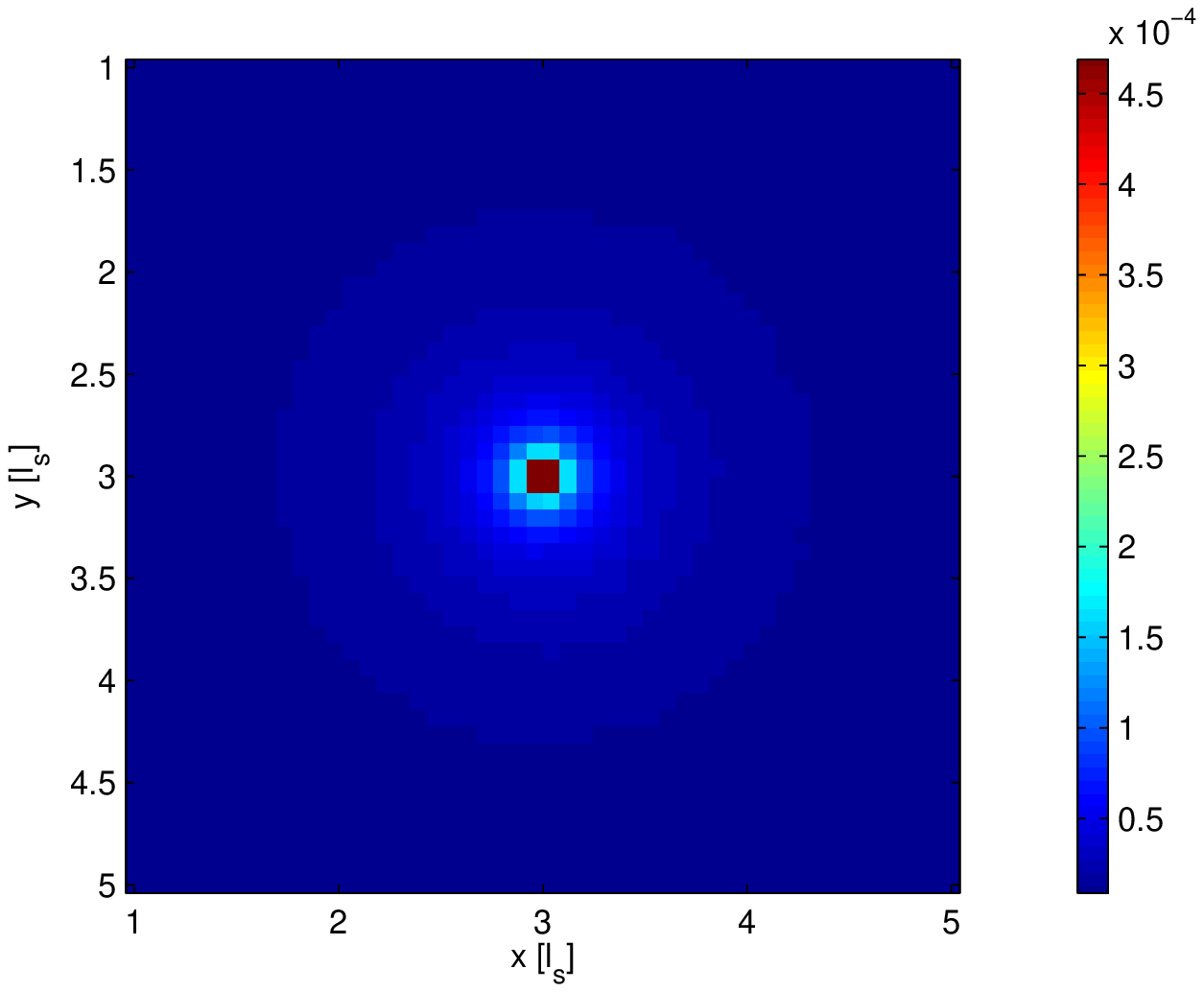}}
\caption{Backscattered continuous-wave intensity for (a) co-polarized and
(b) cross-polarized light with the incident beam.}
\end{figure}

The radial profiles for $g=0.80, 0.85,0.9$ and $0.95$, Fig. (11), as well show no ring formation.
Backscattering is dominated at the point of incidence with no ring features.

The absence of ring formation in the continuous wave backscattered light
of either helicity is attributed to
the smearing of the ring structure where time is integrated over all time steps.
Figures 12 and 13 compare the continuous wave profiles (solid green lines) of preserved
and flipped polarizations,
respectively, to snap shots of the time-resolved profiles for various times.
The continuous wave (time-averaged) profiles have been superimposed on the time-resolved profiles
for comparison.

\begin{figure}
\centering
\subfigure[]{\includegraphics[scale = 0.50]{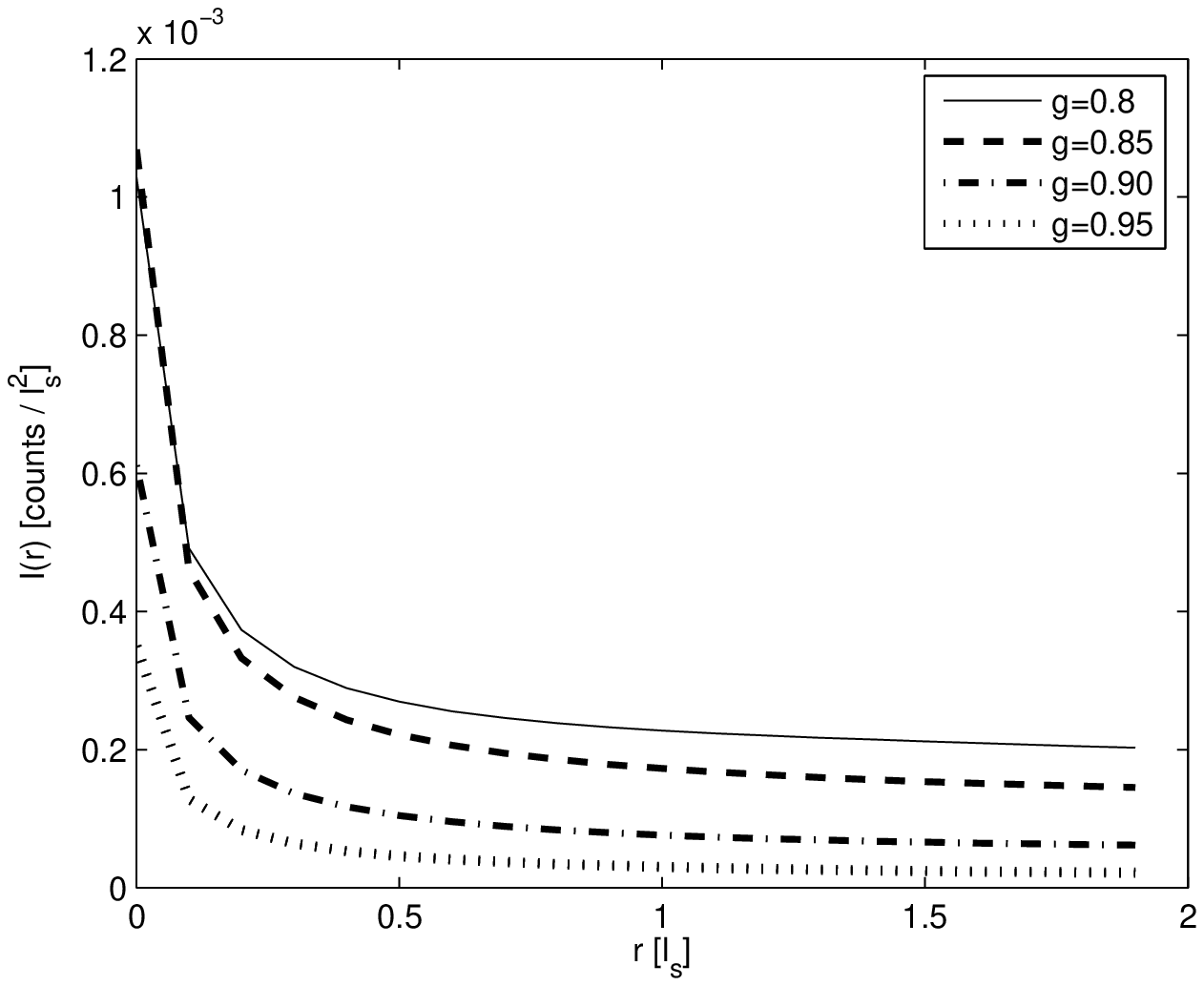}}
\hfill
\subfigure[]{\includegraphics[scale = 0.50]{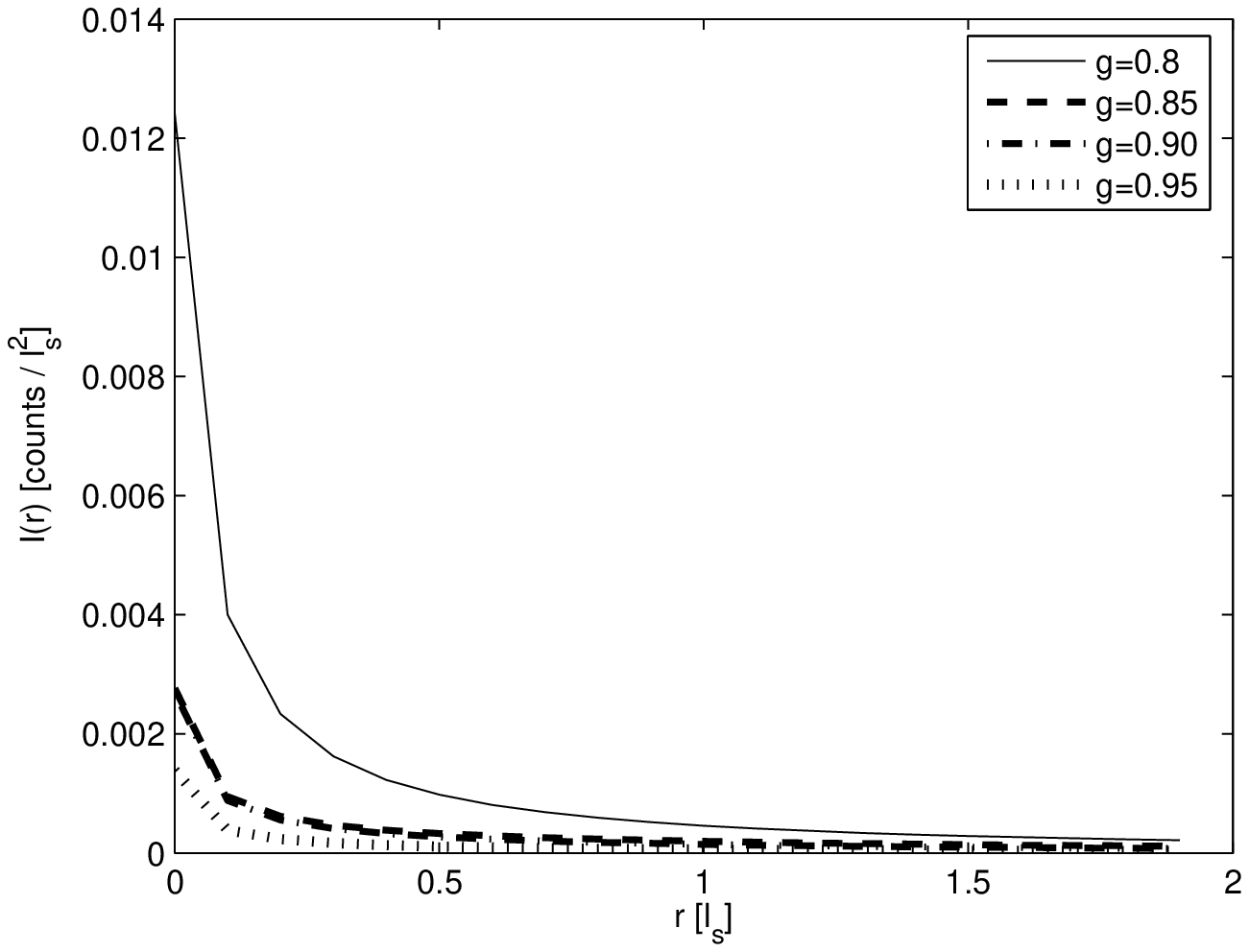}}
\caption{Comparison of backscattered continuous-wave intensity for various anisotropies with the
same (a) and opposite (b) helicity as the incident beam.}
\end{figure}

\begin{figure}
\centering
\includegraphics[scale = 0.75]{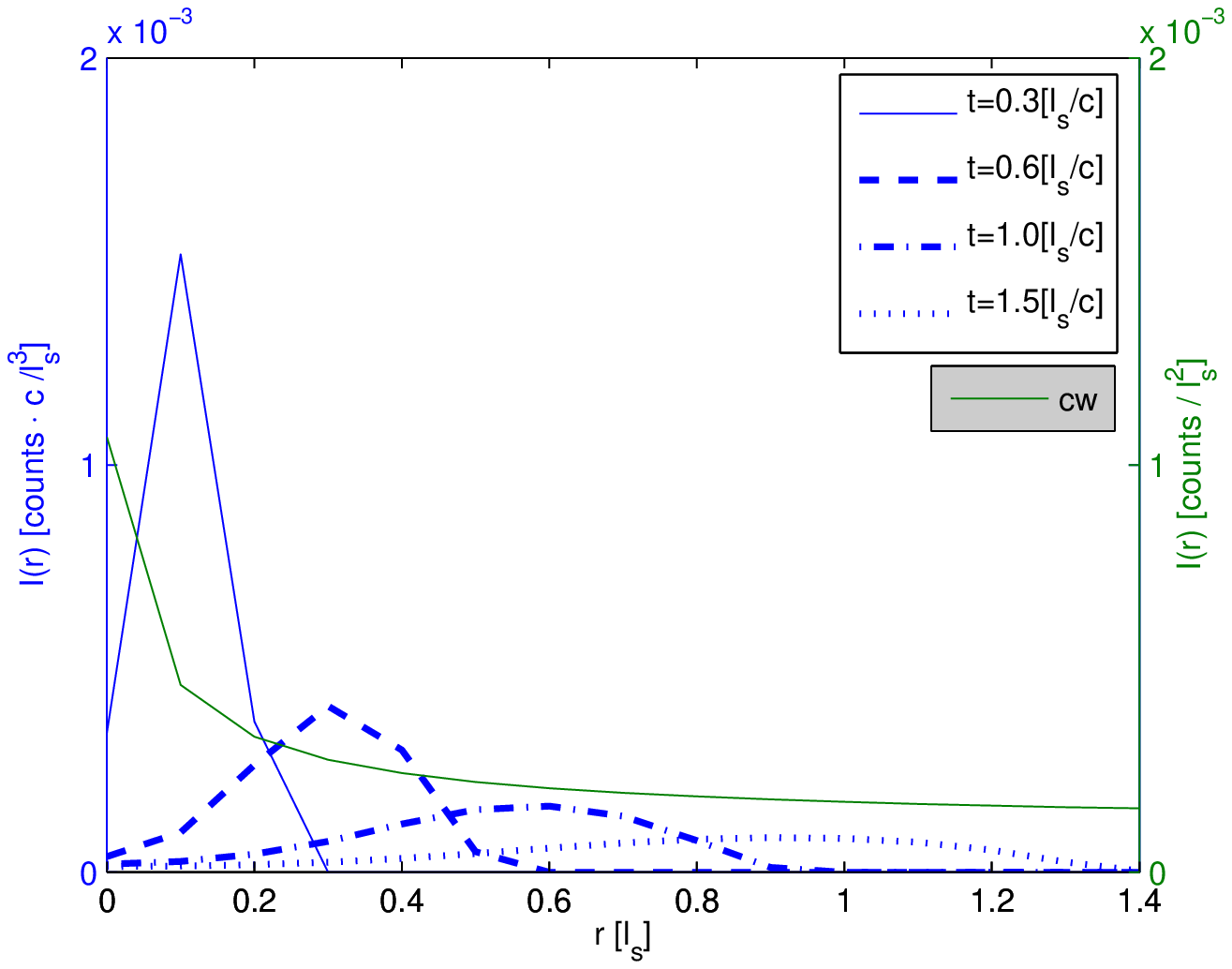}
\caption{Comparison of continuous-wave and time-resolved co-polarized backscattered light for
$g = 0.85$.}
\end{figure}

\begin{figure}
\centering
\includegraphics[scale = 0.75]{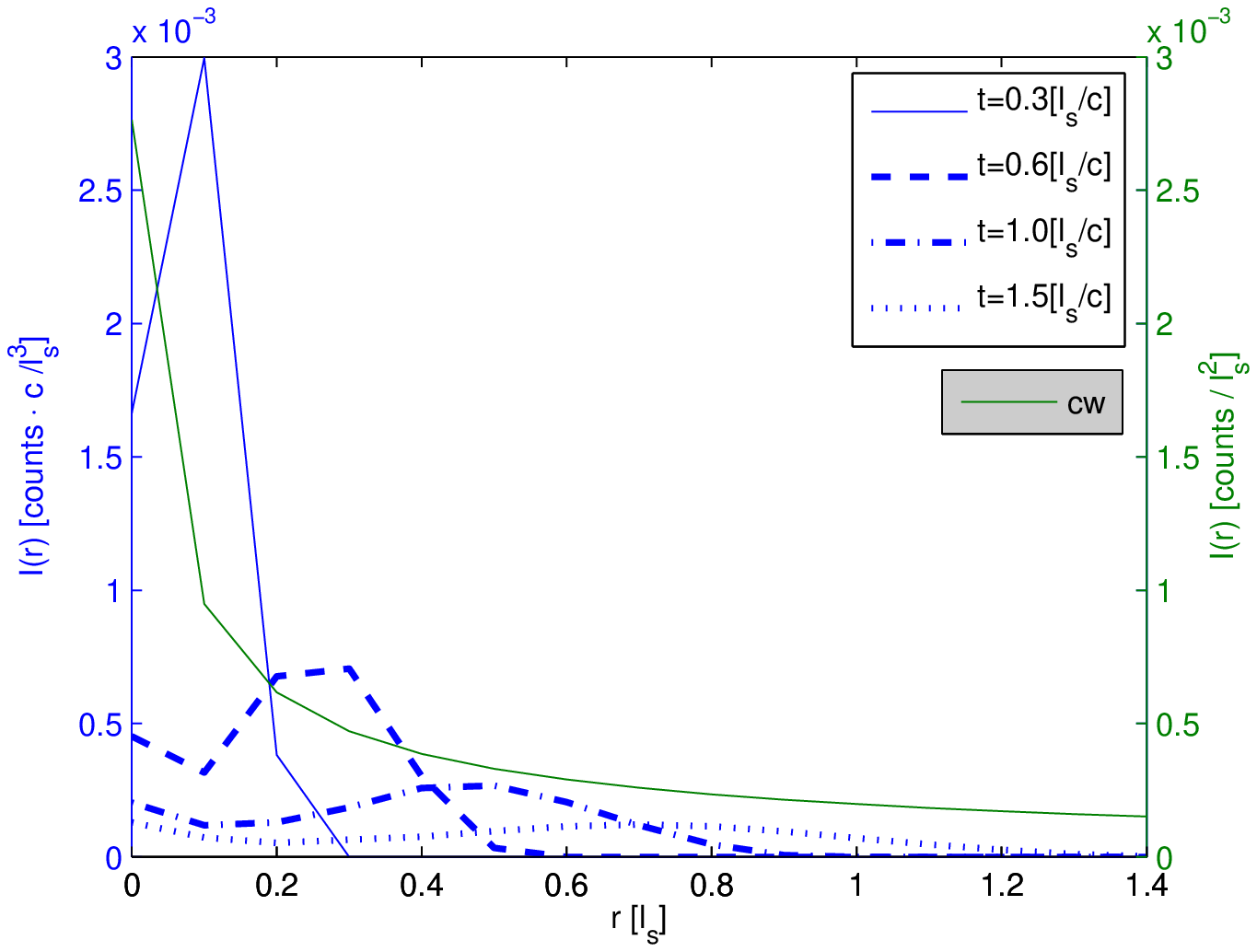}
\caption{Comparison of continuous-wave and time-resolved cross-polarized backscattered light for
$g = 0.85$.}
\end{figure}

\section{Discussion}

Ring formation of backscattered light was first described by Kim and
Moscoso\cite{kim1} who predicted a time-independent ring formation using the scalar time-independent Fokker-Planck
approximation to radiative transfer in forward scattering media. It was also postulated by these
authors that the light
composing the ring was of the same helicity of an incident circularly polarized beam.
Ring formation was described by Kim and Moscoso
as a result of successive near
forward scattering events, due to the forward-peaked scattering nature of the particles, which gave rise to
a steady state semi-circular trajectory of photons inside the medium. The EMC results presented here
suggest partial agreement with Kim and Moscoso
however, EMC differs in that it predicts a
strictly time-\textit{dependent} ring formation. As well, the EMC results predict ring formation for light
backscattered with the same $and$ opposite helicity as that of the incident beam.

The results of the EMC simulation provide a basis for the following
explanation
of backscattered photon transport in forward-peaked scattering media involving time dependence.
In the case of helicity preserved ring formation, the ring peak moves outward away from the point of incidence in time
and grows weak simulatneously. This behavior can be understood in terms of successive near forward scattering events
in which different amounts of light penetrate deeper into the medium than others, see the schematic in Fig. (4).
Light that travels to shallow depths is responsible for ring formation at early times.
Due to the forward-peaked nature of
scattering, light that travels deeper into the medium is predominantly transmitted with only a small constituent
being successively scattered to form arc-like trajectories. This light is responsible for ring formation
at later times.
This gives rise to an ever widening ring radius and
an ever decreasing ring peak as time progresses. As well, it explains why ring peaks are smaller as $g$ increases: light is
scattered about a smaller distribution of angles about the forward direction increasing the
likelihood of transmittance. As well, the arc-like trajectories become longer as anisotropy increases due
to the smaller deviations from
the forward direction of the scattering angle resulting in longer ring lifetimes.
The ring behaviors described here may be experimentally observed using time-resolved
femtosecond pulses
using a streak camera or Kerr gate.

In the case of helicity flipped backscattering, light is backscattered with high likelihood at the point of
incidence and a secondary ring location. Backscattering at the point of incidence is attributed to the structure of the
phase function of the forward-peaked scatterers. There is a finite
peak in the phase function about the scattering angle $\pi$ giving rise to a finite
probability of backscatter with helicity flip.
Because photons are primarily transmitted due to the forward-peaked nature of scattering, light scattered through
a large angle dominates when
analyzing backscatter.
Ring formation in this case is a result of forward-peaked scattering events combined with
large angle scattering events. Forward-peaked scattering brings about the ring behavior
while large angle scattering at some point along the photon's trajectory is responsible for the helicity
flip of the photon and the decreased ring radius, see Fig. (9).
For values of $g < 0.85$, off-forward scattering remains dominant to the point of quenching
opposite helicity ring formation
when collecting backscattered light with arbitrary propagation direction. As a
result, backscatter is dominant at the point of incidence with a Guassian-like distribution as distance from the
central point
of incidence increases.

The analysis of continuous-wave backscatterd light revealed an absence of ring formation. This can be reconciled
with the time-dependent nature of ring formation presented above by recognizing that the time integration
destroys the ring
structure.
It is noted that the dependence of ring formation
on the collection angle of backscattered light is weak in the presence of high
anisotropy, $g \geq 0.85$.
When scattering becomes more isotropic it is expected that ring formation will first be observed when collecting light
exactly backscattered anti-parallel to the incident beam. Lastly, the ring behavior can be observed when collecting light of
both preserved and flipped polarizations simultaneously, Fig. (14).

\begin{figure}
\centering
\includegraphics[scale = 0.75]{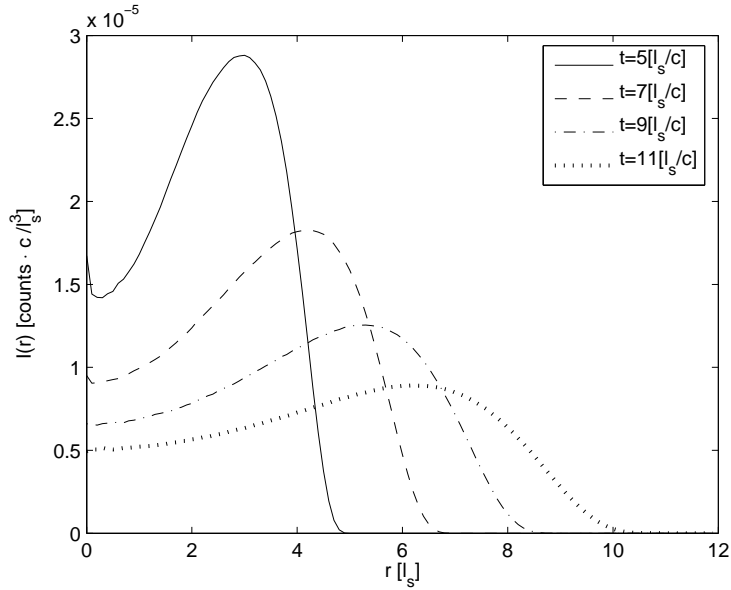}
\caption{Time evolution of simultaneous collection of co-polarized and cross-polarized backscattered light
for $g = 0.85$.}
\end{figure}

\section{Conclusion}

The EMC method was used to investigate the backscattering of circularly polarized light at normal incidence
to a half-space of forward scattering particles ranging in anisotropies of $g = 0.8$ to $g = 0.95$.
The spatial dependence of the backscattered intensity is examined for both the time-resolved and continuous-wave
cases. Time-resolved analysis reveals ring formation for $both$ co-polarized $and$ cross-polarized backscattered
light for $g \geq 0.85$.
For values of $g < 0.85$, off-forward scattering remains dominant to the point of quenching ring formation
when collecting backscattered light with arbitrary propagation direction and opposite polarization with respect to
the incident beam.
Ring behavior is similar in both types of backscattered light: the ring radius grows in time with the
ring-peak decreasing
simultaneously. The ring pattern is more pronounced when light is backscattered with the same polarization as the input.
For the continuous-wave case, no such ring is observed.
The ring formation presented in this study provides an important clue for understanding how light is backscattered from forward scattering media.
Specifically, the EMC results suggest that photons
undergo successive near-forward scattering events.
Futhermore, these findings suggest that backscattered light is comprised of photons undergoing
arc-like trajectories within the
medium. In addition, the photons penetrate to different depths such that rings that
form near the point of incidence of the pencil-like beam (at early times)
are comprised of
photons that have penetrated to smaller depths while rings that form further away from the point of incidence
(at later times)
are comprised of deeper penetrating
photons.
This knowledge may have potential use in polarization imaging techniques to aquire depth information
of targets.
Ring formation with cross-polarized backscattered light arises from
succesive forward-peaked scattering
events over most of the photons path with large angle scattering events taking place at some point along the
trajectory to flip the
photon's helicity. It is the large angle scattering events which give rise to smaller ring radii than in the
case of light backscattered with preserved polarization.
It is noted that these results come in contrast to previous theoretical studies of backscattered light
using the scalar time-independent
Fokker-Planck approximation to radiative transfer\cite{kim1} which predict a continuous-wave ring formation. The results
given by the EMC method suggest that the time dependence and polarization state of the scattered light intensities play a
crucial role in understanding
ring formation for forward scattering media.

\begin{center}
\textit{V. Aknowledgements}
\end{center}

The authors thank  Dr. Florain Lengyel, Assitant Director for Research Computing at the CUNY-GC computing 
facilities, for extensive help and support using the CUNY-GC research cluster. 
The authors acknowledge helpful discussions with
W. Cai. This research was supported in part by ONR Award No: N00014-03-1-0463 and by NASA COSI. Kevin Phillips'
email address is \texttt{kevphill@sci.ccny.cuny.edu}.

\end{document}